\documentclass[a4paper,11pt]{article}
\usepackage{jinstpub} 
\usepackage{lineno}
\usepackage{siunitx}
\usepackage{subfigure}
\usepackage{isotope}

\newcommand*\chem[1]{\ensuremath{\mathrm{#1}}}

\title{Single photon performance characterisation of a Generation I Large Area Picosecond PhotoDetector}

\author[1]{R. J. Foster\note{Corresponding author.},}
\author{A. Scarff}
\author{and M. Malek}
\affiliation{Department of Physics \& Astronomy, University of Sheffield,\\
Sheffield, S10 2TN, United Kingdom}

\emailAdd{r.j.foster@sheffield.ac.uk}

\abstract{The single photoelectron performance characteristics of a Large Area Picosecond PhotoDetector (LAPPD) were studied using a picosecond optical laser source. We verify that the LAPPD is capable of achieving transit time spreads of less than \SI{70}{\pico\second} and spatial resolutions on the order of millimetres when illuminated with single photons. Isolation of the single photoelectron peak is easily possible in the pulse height distribution and the mean single photoelectron gain is measured to be greater than $10^{6}$ with gains above $10^{7}$ possible in certain voltage configurations. We judge the performance of the LAPPD sufficient for application in large-scale water or scintillator-based neutrino experiments, with additional work required to develop the digital signal processing algorithms necessary for processing photon hits in high photon occupancy environments.}


\arxivnumber{1234.56789} 

\begin{document}
\maketitle
\flushbottom

\section{Introduction}
\label{sec:introduction}

The Large Area Picosecond PhotoDetector (LAPPD\textsuperscript{TM}) is a novel microchannel plate (MCP) photosensor that offers single photon timing resolution of tens of picoseconds as well as millimetre-scale spatial resolution \cite{lyashenko2020}.
This performance is achieved in conjunction with a large photosensitive surface, resulting in a relatively low cost per unit area --- an attractive proposition for many applications ranging from high energy physics to medical imaging.
One area in which LAPPDs are of particular interest is in future neutrino experiments \cite{snowmass} such as DUNE \cite{dune}, WATCHMAN \cite{watchman}, and THEIA \cite{theia}.
At Fermilab, the ANNIE experiment \cite{annie} has already successfully deployed LAPPDs inside a water Cherenkov detector \cite{AscencioSosa2024}.
In these experiments, LAPPDs are expected to greatly improve the vertex reconstruction ability and the fast timing is expected to enable new techniques such as Cherenkov-scintillation separation \cite{Caravaca2017,Kaptanoglu2022} in neutrino detectors utilising novel hybrid detection media such as water-based liquid scintillator \cite{Caravaca2020}.

The LAPPD was developed by the LAPPD collaboration, who sought to develop high performance, large area photosensors in order to extract as much information as possible from events at the Fermilab TeVatron \cite{adams2016}.
The LAPPD is manufactured by Incom, Inc., a Massachusetts-based company specialised in the fabrication of glass optical devices.
Traditional MCP photosensors are fabricated using lead glass as both the substrate and the secondary emissive surface \cite{Wiza1979}.
However, there have long been concerns about limits on the total charge that can be extracted from the plate, as well as the complex and expensive fabrication process that results in high costs per unit of photosensitive area.
Conversely, the LAPPD is constructed from inexpensive but robust borosilicate glass which is then later functionalised by atomic layer deposition of a secondary emissive coating \cite{Minot2015}.
This not only enables larger photosensitive areas by reducing the manufacturing cost, but also allows for the electrical properties of the resistive and emissive layers to be tuned depending on the application \cite{Elam2013}.
Since the secondary emissive layer is now decoupled from the substrate, materials with higher secondary emission yields, such as \chem{MgO} or \chem{Al_2O_3} \cite{Insepov2010}, can be used to increase the achievable gain.
In addition, the use of borosilicate glass is advantageous as low radioactivity glass samples can be chosen \cite{Popecki2016} which have far lower \isotope[40]{K} content than lead glass.
Since the presence of \isotope[40]{K} within the glass substrate is a dominant source of background events in MCPs \cite{Siegmund1988}, this dramatically reduces the overall background rate of the MCP.

In this work we aim to verify the single photoelectron performance characteristics of the LAPPD, in particular the manufacturer specifications of transit time spreads below \SI{100}{\pico\second} and gains greater than $10^{6}$.

\section{LAPPD 104}
\label{LAPPD104}

\begin{figure}
    \centering
    \includegraphics[width=0.6\textwidth]{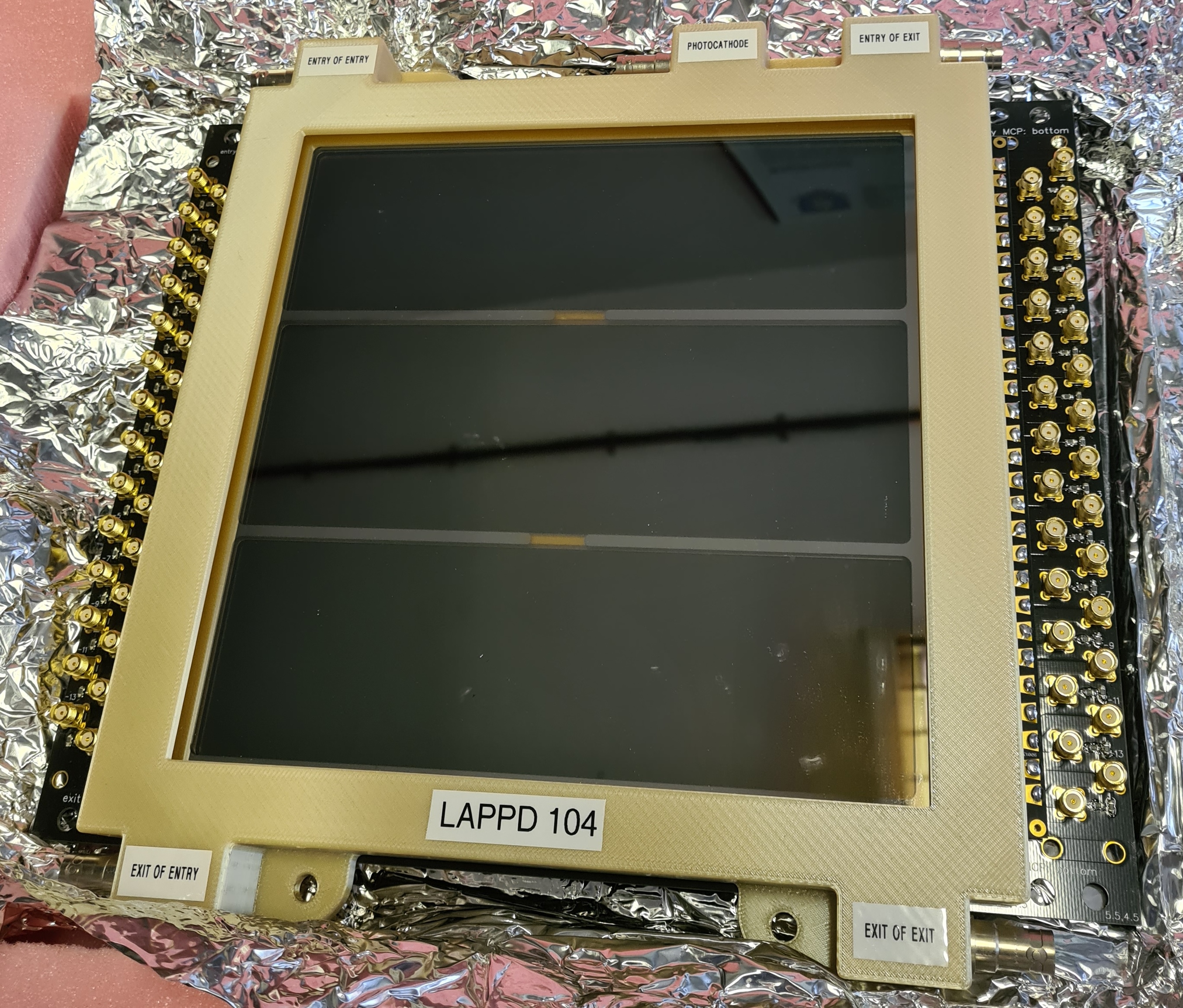}
    \caption{The LAPPD used in this work shown in its carry case. The SMA connectors for signal readout can be seen on the printed circuit boards attached onto either side of the LAPPD.}
    \label{fig:lappd}
\end{figure}

The device characterised in this work was LAPPD 104, a Generation I LAPPD with stripline anode readout, as shown in figure~\ref{fig:lappd}.
LAPPD 104 was manufactured and vacuum sealed in June 2021 and delivered to the University of Sheffield in October 2021.
LAPPD 104 has a \SI{5}{\milli\metre} thick fused-silica glass window.
On the underside of this window the multi-alkali \chem{K_2NaSb} photocathode is deposited, with an average quantum efficiency of 24.3\% at \SI{365}{\nano\metre}.
The active area of the detector is approximately $\SI{195}{\milli\metre} \times \SI{195}{\milli\metre}$.

\begin{figure}
    \centering
    \includegraphics[width=\textwidth]{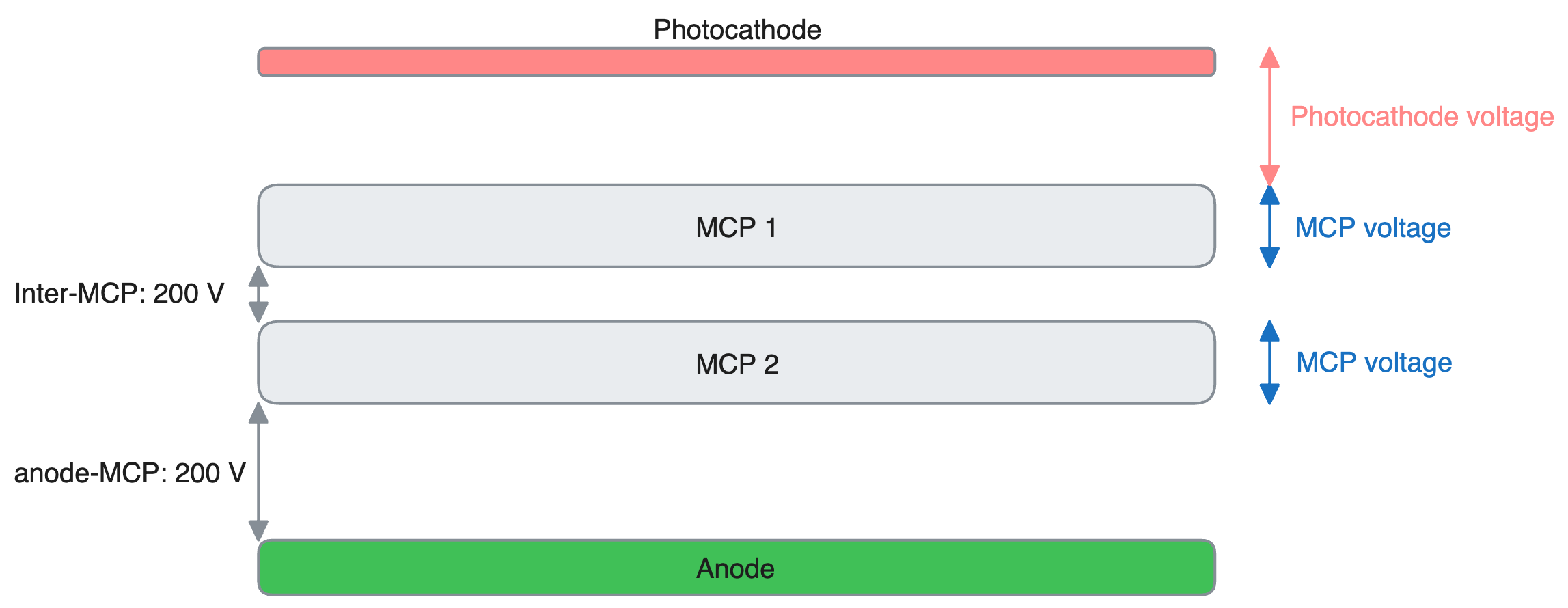}
    \caption{The high voltage configuration for LAPPD 104 during the characterisation measurements. The photocathode voltage and MCP voltage are varied during the tests whilst the inter-MCP and anode-MCP voltage are fixed at \SI{200}{\volt}. The voltage of both MCPs is kept identical to ensure stable operation.}
    \label{fig:voltages}
\end{figure}

LAPPDs contain two MCPs arranged in a chevron pair.
The microchannel pore diameter is \SI{20}{\micro\metre} with a length to diameter ratio of 60:1.
For this device, the secondary emissive layer applied to the MCPs is made from \chem{MgO}.
Five high voltage supplies are required to power the LAPPD - these are applied to the photocathode, and the top and bottom of each of the MCPs in the pair.
In these measurements the voltage between the anode and the bottom of the bottom MCP, and the voltage applied to the gap between the two MCPs, is fixed at \SI{200}{\volt}.
The voltage applied to the photocathode, and the voltage applied across the MCPs themselves, is adjusted as part of the characterisation measurements to determine the optimal operating conditions.
The high voltage configuration is shown in figure~\ref{fig:voltages}.

Charge is collected onto an anode comprising 28 silver striplines which are routed onto a printed circuit board into SMA connectors at both ends of the stripline.
Due to channel constraints on the digitising electronics, double-ended readout of eight striplines was used for the measurements presented here.
Any unused channels were terminated at the SMA connector using \SI{50}{\ohm} terminators.

\section{Test stand setup}
\label{sec:setup}

\begin{figure}
    \centering
    \includegraphics[width=0.75\textwidth]{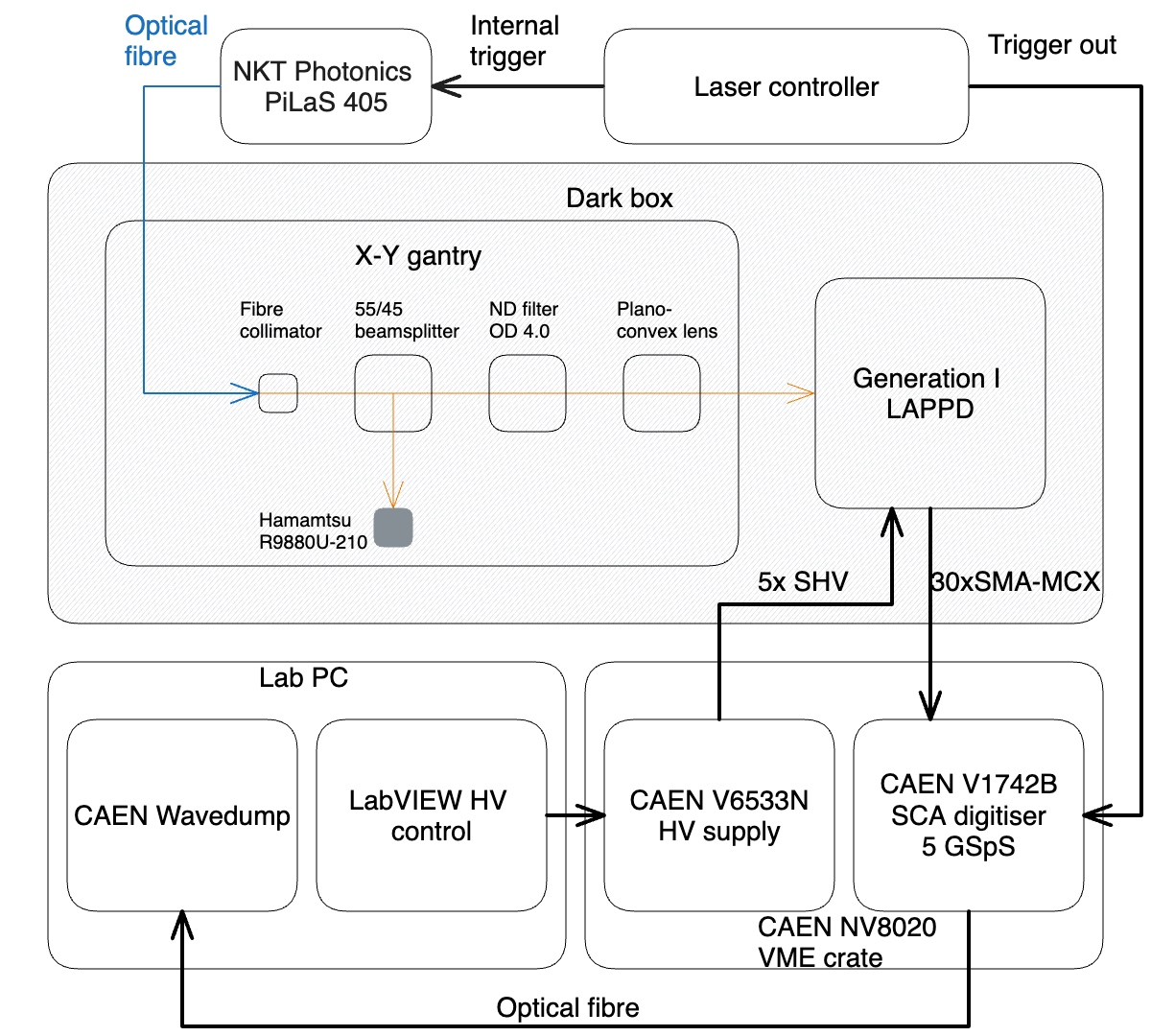}
    \caption{Schematic of the laser test stand setup and associated high voltage and signal digitisation electronics.}
    \label{fig:setup}
\end{figure}

The characterisation measurements were performed using an optical laser test stand.
A schematic of the setup is shown in figure~\ref{fig:setup}.
The LAPPD is mounted onto a vertical plate inside of a large dark box.
It is illuminated using an NKT photonics PILAS DX PIL040-FC picosecond pulsed-diode laser.
The pulse width of the laser is less than \SI{45}{\pico\second}, short enough such that it is possible to make an estimate of the transit time spread of the LAPPD.
The emission wavelength of the laser is \SI{405}{\nano\metre} (spectral width less than \SI{5}{\nano\metre}) which is closely matched with the peak quantum efficiency of the LAPPD bialkali photocathode.
The laser pulse is guided to the LAPPD through a single-mode optical fibre into a fibre collimator coupled to a pellicle beamsplitter, then into neutral density filters, and finally through a plano-convex lens to focus the light onto the photocathode surface.
The total optical density of the neutral density filters was increased up to a value of 4.0, which was sufficient to attenuate the laser light such that the LAPPD only responded to between 6-8\% of laser pulses.
The result is that the vast majority (>95\%) of laser pulses which elicit a response in the LAPPD produce a single photoelectron.
The optical apparatus is attached to an X-Y gantry allowing the laser to be scanned across the LAPPD photocathode.
In this case, the laser was centred in line with a single strip and remained there for all measurements taken.

The five required negative polarity high voltage inputs are provided by a CAEN V6533N VME high voltage power supply.
The high voltage is controlled by a LabVIEW interface allowing for independent control of every channel.
Signal digitisation is performed by a CAEN V1742B switched-capacitor array VME digitiser with a sampling rate of \SI{5}{GS\per\second}.
CAEN WaveDump is used to transfer the digitised signals to a PC over a CONET optical fibre link.

The laser is internally triggered using the laser controller unit at a repetition rate of \SI{100}{\hertz}.
Upon generating a signal ordering the laser to fire, this signal is also propagated to the digitiser ``fast trigger'' input to begin an acquisition.
A total of 16 signal channels are digitised at a time, corresponding to each end of 8 LAPPD striplines.
The trigger signal is also digitised, enabling digital signal processing techniques to be used post data acquisition to form precise relative timestamps between trigger and LAPPD signals.
The trigger signal jitter is <\SI{3}{\pico\second} RMS according to manufacturer specifications.
As a result of the cable delays and delays within the triggering circuitry of the digitiser, there is an arbitrary delay between the trigger signal and the arrival of the laser light at the photocathode.

\section{Signal processing}
\label{sec:processing}

The binary waveform files are decoded and pre-processed before entering the analysis chain for a given measurement.
Firstly, the waveform is baseline-subtracted to remove the arbitrary DC offset by taking the median value of all samples in the waveform before the arrival of the trigger signal.
This value is subtracted from every sample in the waveform.
All values are then converted from ADC counts into millivolts.
The digitiser has 12 bits of resolution and a full-scale range of \SI{1}{\volt} and so each ADC count below baseline corresponded to \SI{0.24}{\milli\volt}.
We intentionally do not apply noise-reducing waveform pre-processing measures such as low-pass filtering to ensure that we preserve as much of the high frequency components in the fast rising LAPPD pulses as possible.

For each event, a region of interest is defined around the likely location of the LAPPD pulse in response to the laser.
The location of this region of interest was determined by measuring the average delay between the arrival of the trigger pulse from the laser and the LAPPD pulse using an oscilloscope.
A \SI{10}{\nano\second} window defines the region of interest.
The integral of the waveform across the window is calculated using trapezoidal integration, which will later be used in the gain measurement.
The maximum amplitude within the region of interest is also calculated and if this value is above \SI{4}{\milli\volt} then a software-implemented constant fraction discriminator is applied to determine the arrival time of the LAPPD pulse.
The arrival time is linearly interpolated in the likely event that the threshold crossing occurs between samples, enabling timestamping precision better than the sampling period of the digitiser.
The constant fraction discriminator is also applied to timestamp the trigger pulse.

\section{Gain}
\label{sec:gain}

\begin{figure}
    \centering
    \includegraphics[width=0.75\textwidth]{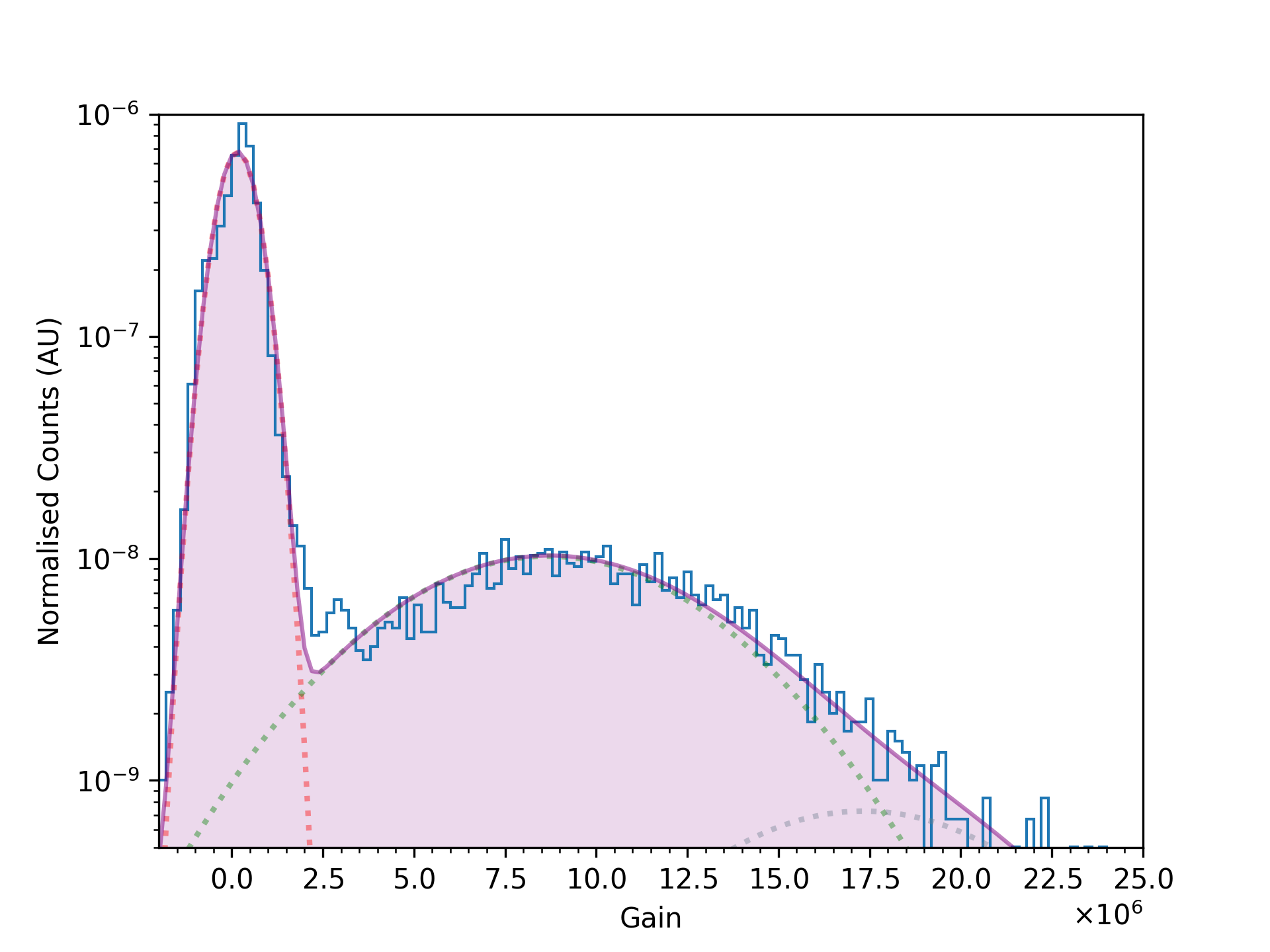}
    \caption{An example of an LAPPD gain distribution fitted with the summation (filled in purple) of three individual Gaussian fits (shown in dashed lines). The first Gaussian represents the charge pedestal, the second represents the single photoelectron charge, and the third is the two photoelectron charge which is largely suppressed due to the low intensity of the incident laser light.}
    \label{fig:fitted_gain}
\end{figure}

\begin{figure}
\centering
\subfigure{\includegraphics[width=0.45\textwidth,keepaspectratio]{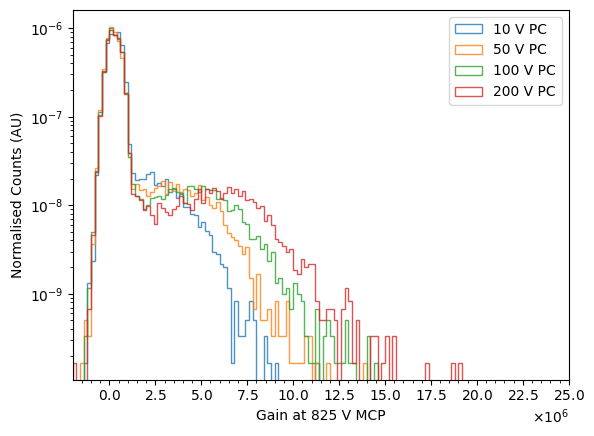}}
\subfigure{\includegraphics[width=0.45\textwidth,keepaspectratio]{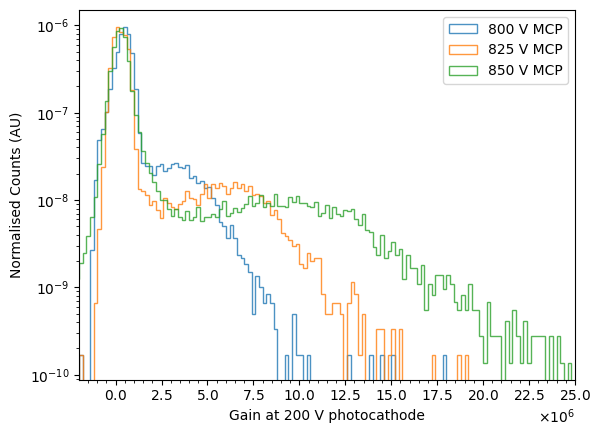}}
\caption{Left: The LAPPD gain distribution with \SI{825}{\volt} applied to the MCPs and varying voltages applied to the photocathode (PC). Right: The LAPPD gain distribution with \SI{200}{\volt} applied to the photocathode and varying voltages applied to the MCPs.}
\label{fig:gain}
\end{figure}

The gain is measured per event by taking the numerical integration of the LAPPD pulse to calculate a charge, taking the termination resistance to equal \SI{50}{\ohm}.
The ratio of the calculated charge to the elementary charge gives the dimensionless value that is then collected into a histogram as is shown in figure~\ref{fig:gain}.
Since most laser pulses do not result in a response from the LAPPD, a large peak representing the charge pedestal can be seen as the first peak in the gain distribution centred around zero gain.
The distribution is fitted with the sum of three Gaussians, the first representing the pedestal, the second representing the single photoelectron peak, and the third representing the two photoelectron peak.
The fitting is constrained such that the mean of the two photoelectron Gaussian must be exactly double that of the single photoelectron peak.
The mean of the single photoelectron Gaussian gives the single photoelectron gain.
An example of this fitting can be seen in figure~\ref{fig:fitted_gain}.

As expected, the mean single photoelectron gain increases with both MCP and photocathode voltage.
It is clear that a gain in the mid $10^{6}$ range is easily achievable even at relatively low operating voltages, with gains approaching $10^{7}$ possible as the voltages are increased.
The gain is more sensitive to the MCP voltage, as the photocathode voltage will only affect the energy of the photoelectron at the initial collision, whereas increasing the MCP voltage increases the energy of all secondary electrons in the multiplication process.



\begin{figure}
    \centering
    \includegraphics[width=0.75\textwidth]{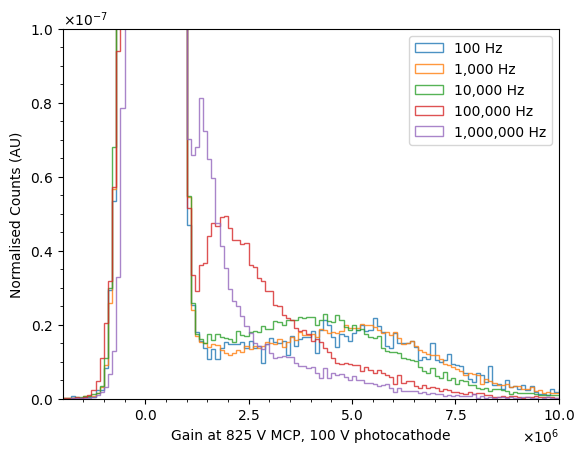}
    \caption{The LAPPD gain distribution as a function of the laser repetition rate, zoomed into the area around the single photoelectron peak to more clearly highlight the differences in the distributions. The laser remains in the same position such that the same microchannels are repeatedly illuminated.}
    \label{fig:laser_rate}
\end{figure}

A microchannel will take a non-negligible time to recharge after performing electron multiplication.
Since the laser light is highly collimated, it is repeatedly striking the same subset of microchannels.
If the same microchannel is struck again before it is able to fully replenish the lost charge, it is expected that the observed gain for that even will be reduced.
By gradually increasing the laser repetition rate, this recharge rate can be investigated.
Figure~\ref{fig:laser_rate} shows the gain distributions from \SI{100}{\hertz} to \SI{1}{\mega\hertz}, demonstrating a reduction in the mean gain of a factor of approximately 3 in the most extreme case.
There is no change in the gain distribution from \SI{100}{\hertz} to \SI{1}{\kilo\hertz}, indicating that the recharge rate is sub-millisecond, although the diameter of the laser spot is certainly larger than the diameter of an individual microchannel and as such we are likely probing the recharge rate of a small collection of pores rather than a single microchannel.

\section{Transit time spread}
\label{sec:tts}

The transit time spread represents the fundamental timing resolution of a photodetector.
Both the LAPPD pulse and the laser trigger pulse timestamps are calculated using a constant fraction discriminator.
The time delta between these timestamps is calculated and histogrammed, forming a Gaussian distribution.
The mean of this distribution is given by a combination the LAPPD's overall transit time, as well as any arbitrary delays resulting from cable lengths and the triggering circuitry of the digitiser, both of which have not been characterised in this study.
The standard deviation of the distribution gives the transit time spread.
In general, the distribution will show a tail which comprises late pulses resulting from the deflection of the photoelectron off the first MCP before it enters a pore and begins the amplification process.
The Gaussian fit only covers the ``core'' distribution and excludes this tail.

\begin{figure}
    \centering
    \includegraphics[width=0.75\textwidth]{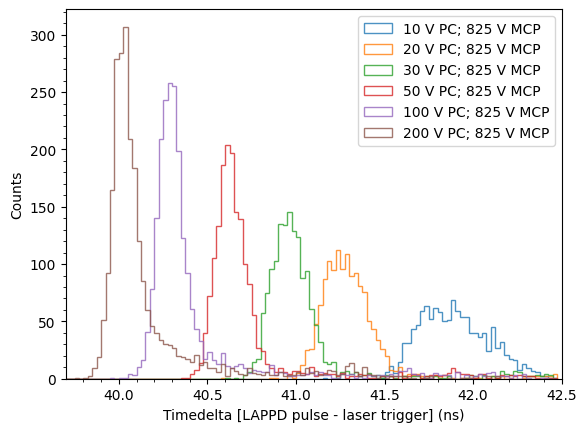}
    \caption{The distributions of the time delta between the laser pulse and the response observed on the LAPPD for varying photocathode voltages.}
    \label{fig:tts_pc}
\end{figure}

\begin{figure}
    \centering
    \includegraphics[width=0.75\textwidth]{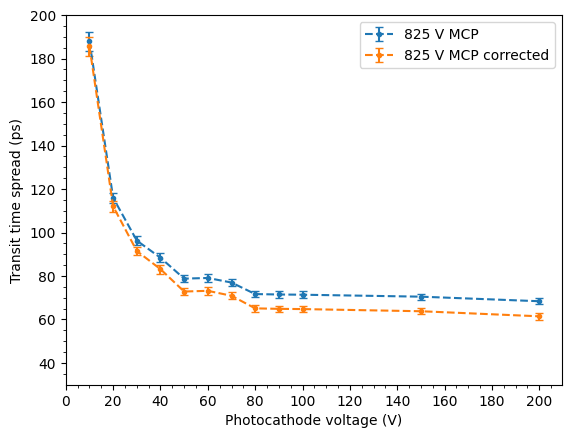}
    \caption{The measured transit time spread as a function of the photocathode voltage with and without correcting for the laser pulse width and trigger jitter.}
    \label{fig:tts_vals_pc}
\end{figure}

From figure~\ref{fig:tts_pc} it is clear to see that increasing the photocathode voltage greatly narrows the observed time delta distribution, resulting in a smaller measured transit time spread.
The overall transit time also decreases, but without suitable characterisation of the aforementioned arbitrary cable and triggering delays it is not possible to calculate an absolute value for the transit time in the current setup.
Figure~\ref{fig:tts_vals_pc} shows the value of the transit time spread as measured by the standard deviation of the Gaussian fit to the time delta distribution across the range of photocathode voltages.
Increasing the photocathode voltage from an initial value of \SI{10}{\volt} to \SI{200}{\volt} results in a reduction of the measured transit time spread from \SI{188}{\pico\second} to \SI{68.4}{\pico\second}.
Adjusting the MCP voltage has very little effect on the resulting transit time spread.
These values are inclusive of the uncertainties arising from the non-trivial pulse width of the laser relative to the transit time spread of the LAPPD.
We can attempt to correct for this by subtracting in quadrature a conservatively small estimate of the laser pulse width of around \SI{30}{\pico\second}, as well as the laser trigger jitter of \SI{3}{\pico\second}.
After this correction, the best transit time spread value we measure is \SI{61.5}{\pico\second} at \SI{200}{\volt} on the photocathode.

\section{Position resolution}
\label{sec:position_resolution}

\begin{figure}[h]
    \centering
    \includegraphics[width=0.9\textwidth]{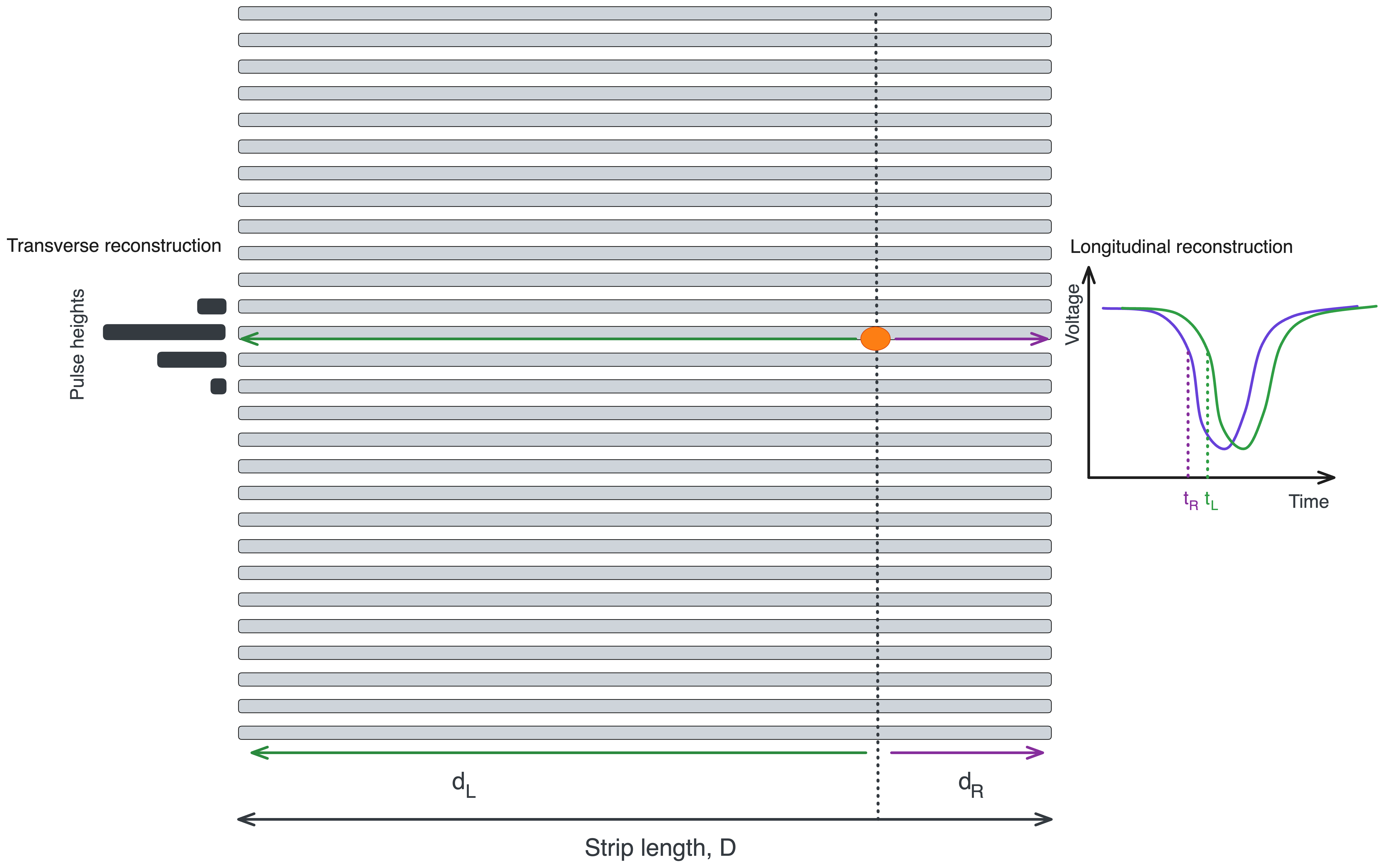}
    \caption{The stripline anode readout and position reconstruction approach on the Generation I LAPPD. The orange dot shows the incident photon location.}
    \label{fig:stripline}
\end{figure}

The photon position is calculated differently in each dimension using the stripline anode readout, shown in figure~\ref{fig:stripline}.
In the direction parallel to the striplines, the photon position is inferred through the difference in the time of arrival at each end of the stripline.
Position resolution in this direction is ultimately dependent on the timing resolution that is achieved.
In the direction perpendicular to the striplines, the photon position is inferred using the relative charge deposition on neighbouring striplines.
The electron cloud will spread as it exits the lower MCP and induce charge in several striplines; the centroid of this charge distribution gives the reconstructed photon position in this dimension.
The position resolution depends on the density of readout elements (i.e. more striplines will improve position resolution at the cost of increasing readout channel density) as well as the gain and the charge spread (inducing charge in more striplines allows for a greater number of striplines to be used during the centroiding process, but requires higher gain to ensure that each stripline observes a charge above the noise threshold).

\begin{figure}[h]
\centering
\subfigure{\includegraphics[width=0.415\textwidth,keepaspectratio]{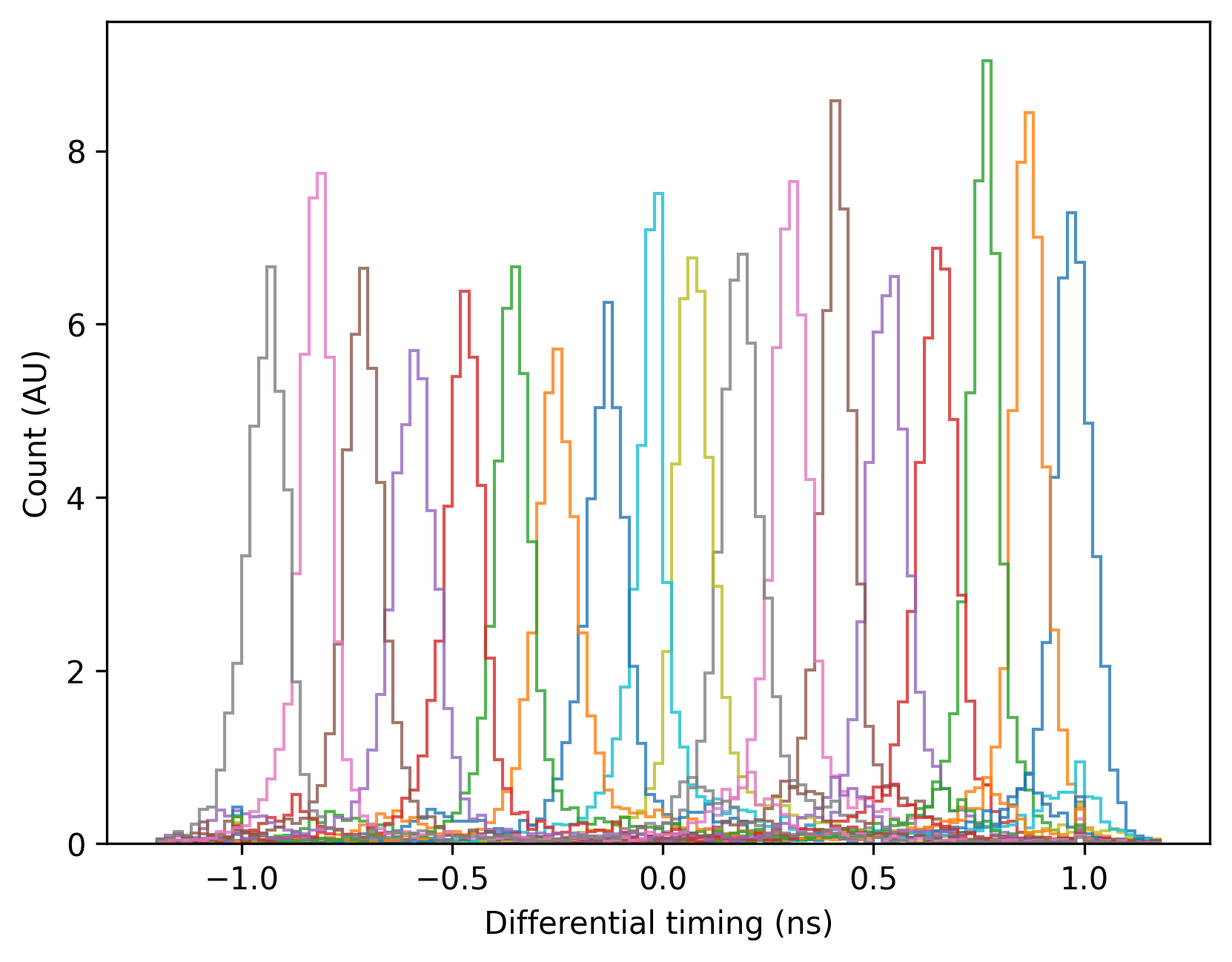}}
\subfigure{\includegraphics[width=0.45\textwidth,keepaspectratio]{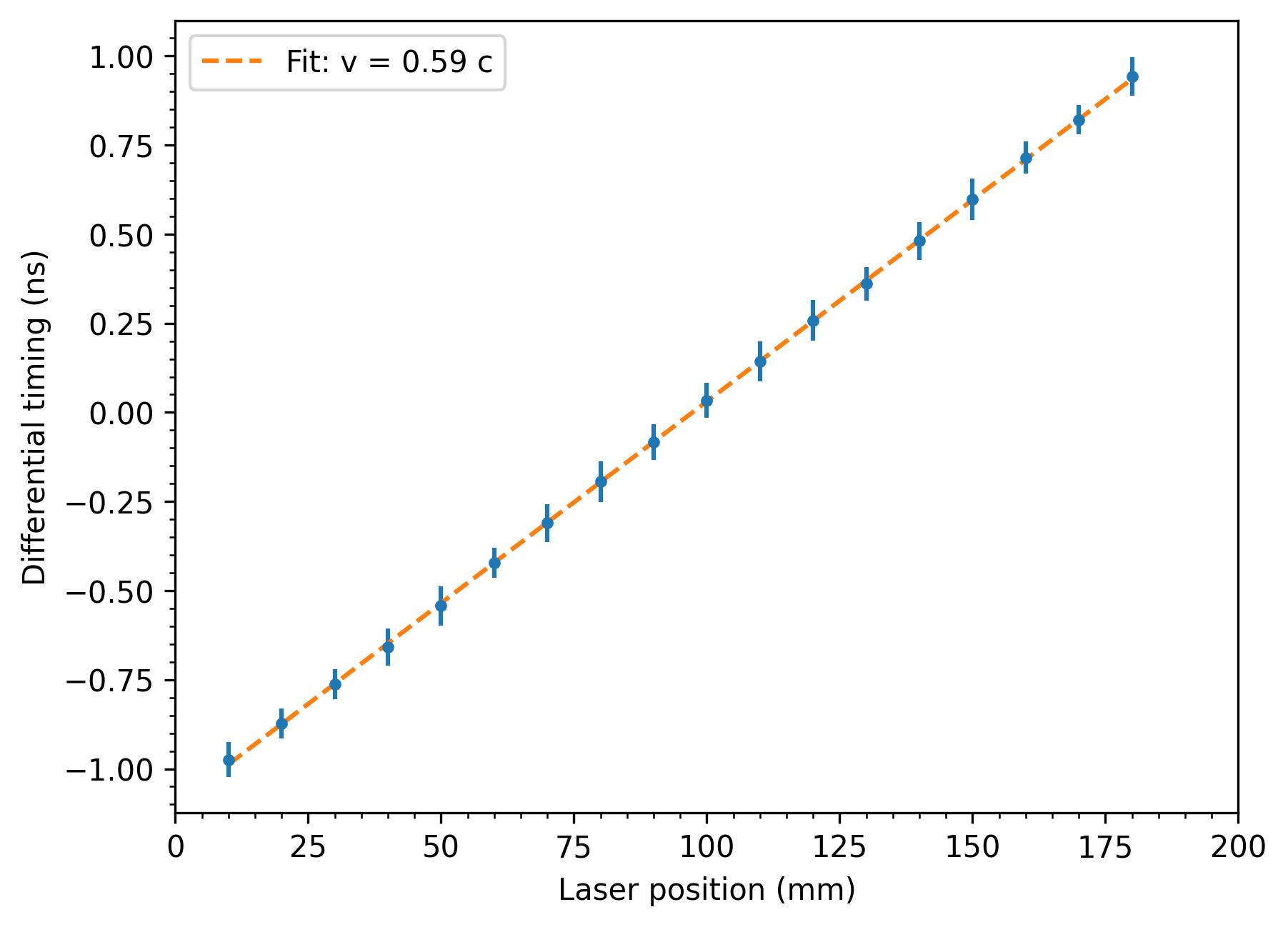}}
\caption{Left: The difference in the pulse time of arrival at each end of the stripline as the laser is scanned along the length of the stripline in \SI{10}{\milli\metre} increments. Right: The mean of each distribution against the laser position. The linear fit gives a stripline pulse propagation velocity of $0.59c$.}
\label{fig:longitudinal}
\end{figure}

\begin{figure}[h]
    \centering
    \includegraphics[width=0.75\textwidth]{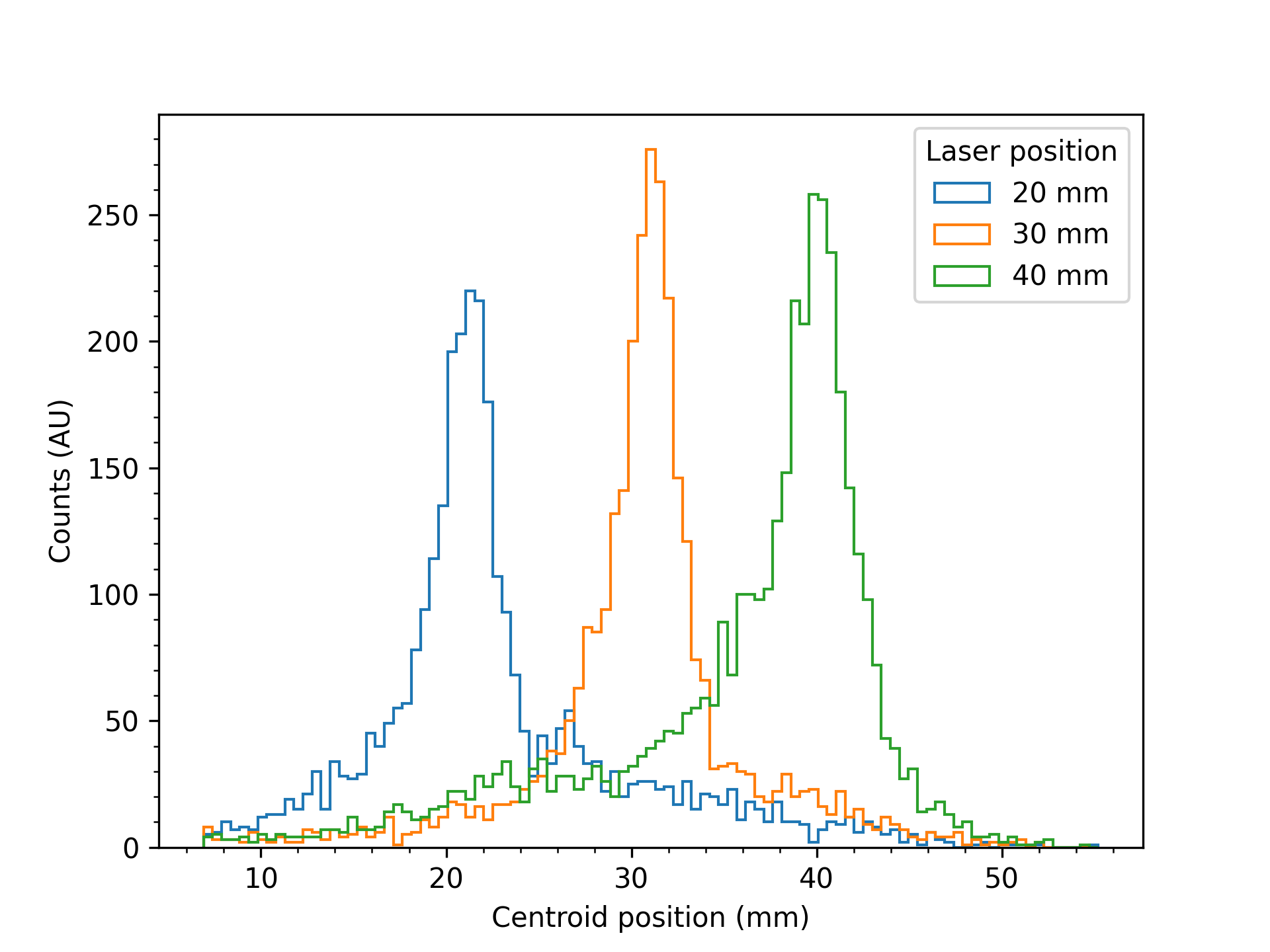}
    \caption{The result of centroiding the charge deposition on 8 striplines for each laser pulse. Three different laser positions are shown, each separated by \SI{10}{\milli\metre} in the direction perpendicular to the striplines.}
    \label{fig:centroid}
\end{figure}

To measure the position resolution in the direction parallel to the striplines, the laser was scanned along the length of a stripline in \SI{10}{\milli\metre} increments.
For each laser pulse, the difference in the time of arrival at each end of the stripline was measured, where constant fraction discrimination with a threshold of 25\% pulse height was used to timestamp each waveform.
The differential timing distributions for each laser position are shown in figure~\ref{fig:longitudinal}.
With an average differential pulse time of arrival resolution of approximately \SI{50}{\pico\second}, this results in a positional resolution in the direction parallel to the striplines of \SI{4.4}{\milli\metre}, which is confirmed when performing the position reconstruction and fitting Gaussian functions to the subsequent position distributions.


Figure~\ref{fig:centroid} shows the result using the charge centroiding approach to infer the photon position in the direction perpendicular to the stripline direction.
The laser is fired at three points, separated by \SI{10}{\milli\metre} in the direction perpendicular to the stripline and the centroid of the charge deposition across 8 striplines is calculated for each laser pulse.
The averaged standard deviation of these distributions, calculated by fitting a Gaussian function to each distribution, across several laser positions gives a position resolution in this dimension of \SI{2.0}{\milli\metre}.

\section{Discussion}
\label{sec:discussion}

We find good agreement between the single photoelectron performance expected from the LAPPD as detailed in earlier characterisation work by Incom \cite{lyashenko2020}, and the performance observed on our laser test stand.
In the case of the transit time spread measurement, the optimal value actually exceeds that measured by Incom, Inc. as part of their standard pre-shipping test and measurement procedure.
For these measurements, the optimal measured value is likely limited by the digitising electronics and signal processing rather than the LAPPD itself, and so these measurements do not represent the maximal performance of the photosensor.
When measuring timing performance in the tens of picoseconds, effects such as the cell-to-cell timing offset of the switched-capacitor array digitiser begin to impact the measurement.
We did not modify the factory digitiser calibration performed by CAEN in this work, although it has been demonstrated that through extensive calibration with signal injection from a precision pulse generator that it is possible to reduce these timing uncertainties to below \SI{1}{\pico\second} \cite{StrickerShaver2014}.

The LAPPD has significant potential for application in high energy physics experiments.
In large-scale water Cherenkov or liquid scintillator neutrino experiments, the position and timing resolution offered by the LAPPD would greatly improve the vertex reconstruction relative to standard photomultiplier tubes which would be hugely beneficial for suppressing backgrounds.
A number of next-generation neutrino experiments are considering instrumenting their detectors with LAPPDs as the single photon performance could unlock a number of proposed goals.
One such goal is the separation of Cherenkov and scintillation light in large-scale water-based liquid scintillator detectors such as the proposed Theia experiment.
Depending on the scintillator used, the LAPPD would be able to consistently resolve the nanosecond-scale time difference between the emission of the prompt Cherenkov and delayed scintillation light.
This would allow for the retention of directional information of the particle by measurement of the Cherenkov ring whilst still benefiting from improved energy and vertex reconstruction arising from the increased light yield from the scintillator.
Cherenkov-scintillation separation has been demonstrated in water-based liquid scintillator using fast photomultiplier tubes \cite{Caravaca2017} and LAPPDs \cite{Kaptanoglu2022} in small-scale or benchtop experiments, but has also been demonstrated in large-scale pure liquid scintillator detectors \cite{borexino}.
The use of LAPPDs in these larger scale experiments would likely allow for improved separation of the Cherenkov and scintillation components as a result of the much better temporal resolution relative to photomultiplier tubes.

There are challenges that must be addressed, however, before LAPPDs are likely to be deployed in any large-scale experiments.
There remains uncertainty about the performance capabilities of the stripline anode readout regarding the ability to disambiguate individual photon hits in high photon occupancy environments.
Previous work \cite{Jocher2016} has shown that deconvolutional approaches show promise in determining the individual hit times and locations on the LAPPD when many photons strike the photocathode in close proximity, however this method has only been tested in simulation.
The High Rate Picosecond PhotoDetector (HRPPD), recently developed by Incom, has the potential to address this issue with a $32 \times 32$ pixel readout array directly coupled to a smaller $10 \times 10$ \si{\centi\metre\squared} photosensitive area which can perform photon counting with millimetre-scale positional resolution without the algorithmic challenges associated with stripline readout.
For experiments which do not require millimetre-scale resolution, the Generation II LAPPD utilises a capacitively-coupled pixellated readout with which it may be possible to count photons on individual pixels.
There have been attempts to develop algorithms for disambiguating photon hits on the stripline anode LAPPD, although none of these have been tested on experimental data so far.
The ANNIE experiment at Fermilab has successfully deployed LAPPDs and a water-based liquid scintillator subvolume recently \cite{AscencioSosa2024}.
Further results are expected in the future which should give an indication of the performance of the stripline anode in high photon density environments.

\acknowledgments

The authors wish to thank Mark Popecki, Cole Hamel, and Stephen Clarke at Incom for their openness and patience in answering questions regarding the LAPPD and for their support in operating the device.
In addition, the authors wish to thank Stephen Wilson and Patrick Stowell at the University of Sheffield for designing the LabVIEW high voltage interface and providing feedback on the draft manuscript, respectively.
This work was supported by the Science and Technology Facilities Council [grant number ST/S006400/1].

\bibliographystyle{JHEP}
\bibliography{biblio.bib}






\end{document}